\documentclass[twocolumn,prl,aps,showpacs]{revtex4}

\usepackage{amsmath}
 \usepackage{bm}
  \usepackage{graphicx}
\usepackage{dcolumn}

\def\dfrac#1#2{{\displaystyle{\frac{#1}{#2}}}}

\begin{document}

\title{Least action principle for envelope functions in abrupt
heterostructures}

\author{A.~V.~Rodina$^{1}$ and A.~Yu.~Alekseev$^{2,3}$}

\affiliation{$^{1}$A.  F.  Ioffe Physico-Technical Institute, 194021,
St.-Petersburg, Russia\\ $^{2}$Department of Mathematics, University
of Geneva, 1211 Geneva, Switzerland\\ $^{3}$Institute of Theoretical
Physics, Uppsala University, S-75108, Uppsala, Sweden } 

\date{\today}

\begin{abstract} 
We apply the envelope function approach to abrupt
heterostructures starting with the least action principle for the
microscopic   wave function.  The interface is treated
nonperturbatively, and our approach is applicable to mismatched
heterostructure.  We obtain the interface connection rules for the
multiband envelope function and the short-range interface terms which
consist of two physically distinct contributions.  The first one
depends only on the structure of the interface, and the second one is
completely determined by the bulk parameters.  We discover new
structure inversion asymmetry terms and new magnetic energy terms
important in spintronic applications.

\end{abstract} 

\maketitle

The envelope function method is a powerful tool which has been widely
used to describe and predict various effects in semiconductors.  It is
normally applicable to  materials with  translation invariance (allowing for
the expansion of the wave function into Bloch functions) and to
slowly varying potentials.  There are two competing approaches to
extending this method to abrupt heterostructures \cite{foreman_prl}
taking into account  interface--related effects.  
 The first one is
to impose appropriate boundary conditions (interface connection rules)
on the envelope wave function at the interface
\cite{bc,gbc,lagr1,ivkamros}.  Another possibility is deriving the
exact envelope function differential equations which are valid near
the interface and which contain the iterface--related terms
\cite{foremanburt,volkov}.  The second approach is more detailed, and
it requires a lot more information on the microscopic structure of the
interface.  Up to now, it has only been applied to
the lattice--matched heterostructures where
the interface is a weak perturbation.  In this case, it has been shown
\cite{foreman_prl} that 
connection rules and differential equations are equally valid
representations of the interface behavior.

It is the aim of this letter to present an extension of the envelope
function method which treats the interface nonperturbatively, and
which is applicable to mismatched heterostructures.  It turns out that
the best approach to the problem is via the Lagrangian variational
principle which encodes the Schr\"odinger equation.  The advantage of
this method is that both the Hamiltonian and the boundary conditions
at the interface are contained in the averaged variational functional.  
The resulting ${\bm k} \cdot {\bm p}$ heterostructure Hamiltonian
coincides with the ordinary ${\bm k} \cdot {\bm p}$ Hamiltonians on
two sides of the interface.  In addition, it contains short--range
interface (SRI) terms which are the main object of our study.  We show
that the SRI terms consist of two physically distinct contributions.
The first one is represented by the Hermitian interface matrix.  Its
components are directly connected to the parameters of boundary
conditions for the envelope functions 
\cite{bc} and determined by the microscopic structure of the
interface.  The second contribution is completely determined by the
bulk parameters of the materials.  
It includes new structure inversion asymmetry (SIA) terms and new SRI
magnetic terms that are additional to the well known Rashba SIA terms
and to the macroscopic magnetic terms, respectively.  Taking them into
account is important for various mechanisms of spin polarization, spin
filtering and spin manipulation that play a key role in semiconductor
spintronic applications \cite{spintronic}.

\begin{figure}
 \begin{center} 
\includegraphics[width=7.5 cm,height=11 cm]{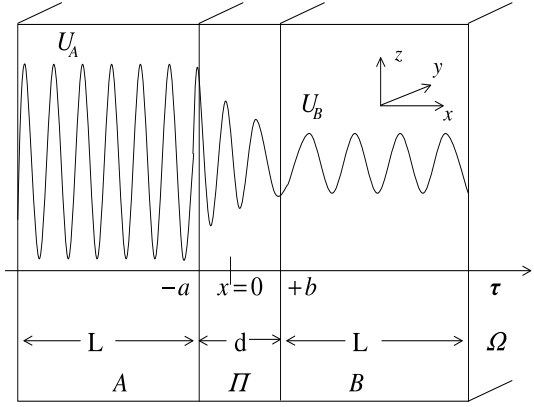}
\vskip -4cm
\end{center} 
\caption{\label{Fig1} Sketch of the planar
heterointerface between $A$ and $B$ semiconductor layers.  $\Pi$
denotes the boundary region. $U_A$($U_B$) is 
the crystal potential in $A$($B$).}
  \end{figure}

In this letter we consider a model of a semiconductor heterostructure
made of two homogeneous semiconductor layers $A$ and $B$ of
characteristic length $L$.  The layers are joined by a thin boundary
region $\Pi$ of the width $d \approx a_0 \ll L$ (see Fig.
\ref{Fig1}), where $a_0$ is the lattice constant.  We work in the
single electron approximation, and we denote by $U(\bm r)$ the
effective potential for electrons.  $U(\bm r)$ coincides with periodic
crystal potentials $U_{A,B}(\bm r)$ inside the bulk--like regions $A$
and $B$, respectively.

We start with the microscopic Lagrangian variational principle which
encodes the stationary single electron Schr\"odinger equation.  The
corresponding Lagrangian density is of the form, 
\begin{eqnarray} 
{\cal L}(\Phi^*,\Phi)&=& (E- U(\bm r))|\Phi({\bm r})|^2 -
\frac{\hbar^2}{2m_0}|\nabla \Phi|^2 \, .  
\label{calL}
 \end{eqnarray}
Here $m_0$ is the free electron mass and $\Phi$ is the microscopic
spinor wave function.  To simplify the presentation we first neglect
the spin-orbit terms in Eq.  (\ref{calL}).  The variational principle
reads, 
\begin{eqnarray} 
\delta {\cal S}&=&\delta \int_\Omega {\bm d}^3
{\bm r} { \cal L}(\Phi^*,\Phi) =0 \, , 
\label{cals} 
\end{eqnarray}
where variations $\delta \Phi$ and $\delta \Phi^*$ are independent of
each other and vanish at the outer boundaries of the integration
region $\Omega = A + \Pi +B$.  The variational principle implies the
microscopic Schr\"{o}dinger equation $\hat H_{\rm micro}{ \Phi}({\bm
r})= E { \Phi}({\bm r})$ with the microscopic Hamiltonian $\hat H_{\rm
micro}= ({\hat p^2}/{2m_0} + U(\bm r))$, where $\hat{\bm p} = -i \hbar
{\bm \nabla}$ is the momentum operator.  The microscopic probability
flux density ${\bm j}=({ \Phi}^*({\bm r}) \hat {\bm p}{ \Phi}({\bm r})
+ \hat {\bm p}^*{ \Phi}^*({\bm r}) { \Phi}({\bm r}))/2m_0$ is
conserved:  ${\bm \nabla} {\bm j}=0$, the microscopic wave function
$\Phi$ is continuous everywhere in the heterostructure.

It is our aim to pass from the description in terms of the microscopic
wave function $\Phi$ to the envelope function approximation.  To this
end, we use expansions ${ \Phi ({\bm r})} = \sum_{n=1}^{N_{A,B}}
\Psi_{n}^{A,B}({\bm r}) u_{n}^{A,B}$ within $A$ and $B$.  Here
$u_{n}^{A,B}$, $n=N_{A,B}$ are the Bloch functions at extremum points
of the bulk energy band structure.  The $N_{A,B}$ component envelope
functions $\Psi_n^{A,B}({\bm r})$ are smooth in the $A$ and $B$
regions where they satisfy matrix Schr\"odinger equations $\hat
H^{A,B} (\hat {\bm k}) { \Psi}^{A,B}({\bm r})= E { \Psi}^{A,B} ({\bm
r})$.  Here $\hat H^{A,B} = \hat C + \hbar \hat { B}^\mu\hat{ k}_\mu +
\hbar^2 \hat D^{\mu \, \nu} \hat k_\mu \hat k_\nu$ are the standard
${\bm k} \cdot {\bm p}$ Hamiltonians including the terms up to the
second order in the wave vector operator $\hat{\bm k} = -i {\bm
\nabla}$.  The matrices $\hat C$, $\hat{B}^{\mu}$ and $\hat D^{\mu \,
\nu}$ ($\mu \, , \nu = x,y,z$) are Hermitian $N_{A,B} \times N_{A,B}$
tensors of rank $0$,$1$, and $2$, respectively.  The Hamiltonians
$\hat H^{A,B}$ give a direct description of the $N_{A,B}$ basic bands
as well as the contributions of the remote bands in the second order
of perturbation theory \cite{bir}.  Note that symmetry of the
materials, the number of basic bands in the ${\bm k} \cdot {\bm p}$
approximation can be different on two sides of the interface.
Moreover, parameters of bulk Hamiltonians
$\hat H^{A,B}$ can vary significantly across the interface,
so as it cannot be treated as a weak perturbation of
the bulk problem.

We fix once and for all the basic functions $u^{A,B}$ in $A$ and $B$,
and we derive the ${\bm k} \cdot {\bm p}$ version of Lagrangian
variational principle.  The result has the form $\delta {\cal S}_{\rm
kp}= \delta {\cal S}_A+\delta {\cal S}_B + \delta {\cal S}_{\rm
sur}=0$, and it contains $\tilde N=N_A+N_B$ independent variations of
the envelope wave functions $\delta \Psi_{n}^{A}, \delta
\Psi_{n}^{B}$.  Here $\delta {\cal S}_{A,B} = \int_{{{A,B}}} {\bm d}^3
{\bm r} { \cal L}({ \Phi}^*,{ \Phi}) = \int_{{{A,B}}} {\bm d}^3 {\bm
r} { \cal L}_{A,B}({ \Psi}^*,{ \Psi})$ and $\delta {\cal S}_{\rm
sur}=\delta \int_{{\bm \Pi}} {\bm d}^3 {\bm r} \delta(x) {\cal L}_{\rm
sur}$, where $\delta(x)$ is the Dirac delta--function.  The bulk
multiband Lagrangian densities ${ \cal L}_{A,B}$ are obtained from the
microscopic Lagrangian by averaging over the unit cells in $A$ and
$B$, respectively.  They have the form:
  \begin{eqnarray}
   && {\cal L}({ \Psi}^*,{ \Psi}) = E |\Psi|^2 -\Psi^* \hat C\Psi -
\frac{\hbar^2}{2} {\nabla}_\mu \Psi^*\hat A^{\,\mu \, \nu}
{\nabla}_\nu \Psi \label{hl} \\
&& - \frac{i\hbar}{2} \left( {\bm
{\nabla}} \Psi^* \hat {\bm B} \Psi - \Psi^* \hat {\bm B} {\bm
{\nabla}} \Psi \right) + \frac{\hbar^2}{2} {\bm \nabla} \Psi^* \cdot
[\hat {\bm K} \times {\bm \nabla} \Psi]\, .  \nonumber 
\end{eqnarray}
Here $\hat A^{\mu \, \nu} = \hat D^{\mu \, \nu} + \hat D^{\nu \,
\mu}$, $\hat K^{\eta} = \epsilon_{\eta \mu \nu } \hat D^{\mu\, \nu}$,
$\eta,\mu, \nu =x,y,z$ and $\epsilon_{xyz}$ is Levi-Civita
anti-symmetric tensor.  The surface Lagrangian is nonlocal and it can
be written as  
\begin{equation} {\cal L}_{\rm sur} =\tilde \Psi^{*}
\hat T_{\rm sur} \tilde \Psi \, , \quad \hat T_{\rm sur} =\frac{
\hbar^2}{2m_0} \left( \begin{array}{cc}\hat t^A /a & \hat t/d \\ \hat
t^{*}/d & \hat t^B /b\end{array} \right) \, , \end{equation}
where
 
$\tilde \Psi = \left( \begin{array}{c} \Psi^A({\bm \rho},x-a) \\
\Psi^B({\bm \rho},x+b)\end{array} \right)$, with ${\bm \rho} = (y,z)$,
and $d=a+b$ (see Fig.  \ref{Fig1}). 
 The energy independent hermitian
$\tilde N \times \tilde N$ interface matrix $\hat T_{\rm sur}$ depends
on the symmetry of both bulk materials and of the interface.  It can
be constructed by using the method of the invariants \cite{bir} or
calculated directly via the microscopic modeling of the potential
$U(r)$ in the interface region $\Pi$ (the details will be presented
elsewhere).

The effective Lagrangians ${\cal L}_{A,B}$ together with ${\cal
L}_{\rm sur}$ contain all the relevant information about the bulk and
interface properties of the heterostructure.  Application of the least
action principle $\delta {\cal S}_{\rm kp}=0$ generates the
Schr\"{o}dinger equation $\hat H_{AB} \tilde \Psi = E \tilde \Psi$
with the complete heterostructure ${\bm k} \cdot {\bm p}$ Hamiltonian
$\hat H_{AB}$ and the general boundary conditions (GBC) to be imposed
on $\Psi^a=\Psi^A({\bm \rho},-a)$ and $\Psi^b=\Psi^B({\bm \rho},b)$.
The GBC (see \cite{bc}) can be written as 
$\left( \begin{array}{c} i
\hat {V}_\tau\Psi^a\\-i \hat {V}_\tau\Psi^b\end{array} \right)
=\dfrac{2\hat T_{\rm sur}}{\hbar} \left(
\begin{array}{c}\Psi^a\\\Psi^b\end{array} \right)$ or as $\left(
\begin{array}{c} \Psi^a\\i \hat {V}_\tau\Psi^a\end{array} \right) =
\hat T_{\rm tr} \left( \begin{array}{c}\Psi^b\\i \hat
{V}_\tau\Psi^b\end{array} \right)$,
 where the components of the
$2N_A\times 2N_B$ transfer matrix $\hat T_{\rm tr}$ (see
\cite{bc,gbc}) can be readily expressed via the components of the
surface matrix $\hat T_{\rm sur}$ (see also \cite{balian}).  Here
${\bm \tau}$ is the normal vector to the interface, $\hat V_\tau={\bm
\tau} \cdot \hat {\bm V}$, and the envelope velocity operator $\hat
{\bm V}_{nm} \Psi_m = {2i}/{\hbar}({\partial {\cal L}}/{\partial {\bm
\nabla }\Psi^*_n}) $ can be written explicitly as 
\begin{eqnarray}
\hat {\bm V} = \hat {\bm B} + \hbar \frac{\partial \hat A^{\mu \nu}\{
\hat k_\mu \hat k_\nu \}}{\partial {\bm k}} - \hbar \left[\hat {\bm K}
\times \hat {\bm k} \right]\, .  
\label{vel} \end{eqnarray} 
The last
term is new in comparison to \cite{bc}.  The corresponding extra term
in the envelope flux density $ {\bm J}({\bm r}) = {1}/{2} \left( {
\Psi}_n^* \hat {\bm V}_{nm}{ \Psi_m} + { \Psi_n} (\hat {\bm V}_{nm}{
\Psi_m})^* \right) $ is proportional to ${\bm \nabla} \times
(\Psi^*\hat {\bm K}\Psi)$ and does not alter the continuity property
${\bm \nabla} \cdot {\bm J}=0$.  It is straightforward to verify that
$J^{\alpha \, \beta}_\tau={\bm \tau} \cdot{\bm J}^{\alpha \,
\beta}({\bm r})={\rm const}$, where $\alpha$ and $\beta$ label two
functions $\Psi_\alpha$ and $\Psi_\beta$ satisfying the same GBC (see
Ref.  \cite{bc}).  This ensures that the heterostructure ${\bm k}
\cdot {\bm p}$ Hamiltonian $\hat H_{AB}$ is self-adjoint.  It has the
form: 
 \begin{eqnarray} \label{HAB} \hat H_{AB}= \left(
\begin{array}{cc} \hat H^A_h + \delta (x) \hat H_{\rm sri}^A & \delta
(x)\hat H_{\rm sri}^{AB} \\ \delta (x) \hat H_{\rm sri}^{BA} & \hat
H^B_h + \delta (x) \hat H_{\rm sri}^B \end{array} \right) \, ,
\label{hab} \end{eqnarray} where $\hat H^{A}_h = \Theta(- x) \hat
H^{A}({\bm \rho},x-a)$, $\hat H^{B}_h = \Theta( x) \hat H^{A}({\bm
\rho},x+b)$, $\Theta (x)$ is the Heaviside step function and
\begin{eqnarray} \hat H_{\rm sri}^A=\frac{\hbar}{2} \left( i\hat
{V}_{\tau}^A + \frac{\hbar}{m_0 a} \hat t_a \right) \, , \quad H_{\rm
sri}^{AB}=\frac{\hbar^2}{2m_0 d} \hat t \, , \\ \hat H_{\rm
sri}^B=\frac{\hbar}{2} \left( -i\hat {V}_{\tau}^B + \frac{\hbar}{m_0
b} \hat t_b \right) \, , \quad H_{\rm sri}^{BA}=\frac{\hbar^2}{2m_0 d}
\hat t^* \, .  \nonumber \label{hsur} \end{eqnarray}
We see that there
are two physically distinct contributions to the short--range
interface (SRI) terms of the Hamiltonian $\hat H_{AB}$.  The first one
arises from the nonlocal surface Lagrangian ${\cal L}_{\rm sur}$ and
it depends on the properties of the interface via the energy
independent parameters of the GBC.  The other contribution comes from
the velocity operator $\hat V_{\tau}$.  It is entirely determined by
the bulk parameters and it arises from the nonvanishing variation of
the bulk Lagrangians ${\cal L}_{A,B}$ at the interface.  The important
feature of this contribution is the presence of the asymmetric $\hat
{\bm K}$ term.  In homogeneous semiconductors the asymmetric $\hat
{\bm K}$ term does not contribute to $\hat H$ in the absence of
external fields (see \cite{lut}).  Examples below demonstrate that the
$\hat {\bm K}$ terms in the Lagrangian of Eq.  (\ref{hl}) and in the
velocity operator of Eq.  (\ref{vel}) become important if the symmetry
is broken due to the presence of external fields or asymmetric
interfaces.  To emphasize this point we neglect effects of bulk
inversion asymmetry.

As a first example, we consider the effective mass Hamiltonian $\hat
H(\Gamma_6) = E_c + {\hbar^2 \hat k^2}/{2m}$ for the spinor envelope
function $\Psi_\alpha$ ($\alpha=\pm 1/2$), where $E_c$ is the bottom
of the conduction band and $m$ is the effective mass.  Following our
method we introduce the effective mass Lagrangian density
\begin{equation} {\cal L} = (E-E_c)|\Psi|^2 -
\frac{\hbar^2}{2m}|\nabla \Psi ({\bm r})|^2 + {\cal L}_{\rm SIA} \, ,
\label{lema} \end{equation} 
which contains the asymmetric term
\begin{eqnarray}
{\cal L}_{\rm SIA}(\Gamma_6) = - \frac{i\hbar^2}{4m_0}
\tilde g {\bm \nabla} \Psi^* [{\bm \sigma} \times {\bm \nabla} \Psi]
\, \label{lesia} \end{eqnarray}
 obtained with ${\bm K} (\Gamma_6) =-(
i\tilde g/2m_0 ){\bm \sigma}$, where $\sigma_x$, $\sigma_y$,
$\sigma_z$ are Pauli matrices, and $\tilde g =g_0 -g$ is the
difference between free electron and effective electron $g$ factors.
Note that $\tilde g \ne 0$ only if the spin-orbit splitting $\Delta$
of the valence band or $\Delta^c$ of the remote conduction band is
taken into account.  We discover now that it is this asymmetric term
${\cal L}_{\rm SIA}(\Gamma_6)$ (missing in Refs.  \cite{lagr1,lagr})
that induces the spin dependence of the velocity operator $\hat {\bm
V}=( \hbar/m )\hat {\bm k}- ({i \hbar} \tilde g/2{m_0})[ {\bm \sigma}
\times\hat {\bm k} ] $ and thus the spin dependence of the standard
boundary conditions $ \Psi={\rm const}$, $\hat V_\tau \Psi= {\rm
const}$ at the interface (see \cite{vasko,pfeffer}).  The short-range
interface SIA term in the heterostructure Hamiltonian $\hat H_{AB}$ of
Eq.  (\ref{hab}) also results from ${\cal L}_{\rm SIA}(\Gamma_6)$.
Moreover, exactly this term ${\cal L}_{\rm SIA}(\Gamma_6)$ produces
the magnetic energy term $-{1}/{2} \mu _{B}\tilde g({\bm \sigma}{\bm
H})$ additional to the free electron magnetic energy ${1}/{2} \mu
_{B}g_{0} ( {\bm \sigma}\, {\bm H})$ in the bulk semiconductor in
external magnetic field ${\bm H}$.  Here $ \mu _{B}$ is the Bohr
magneton.   
Next, it can be shown that the macroscopic SIA term
$\hat H^{\rm so}=\alpha_R [ {\bm \sigma} \times\hat {\bm k} ] {\bm
\tau}$, postulated by Rashba \cite{rashba} for the asymmetric $2D$
structure, is generated by the term ${\cal L}_{\rm SIA}(\Gamma_6)$.
For this the dependence of $g$ on the potential $V = -|e|{\cal E }x$
should be taken into consideration, where the average electric field
${\bm E}= {\cal E} {\bm \tau}$ characterizes the macroscopic
asymmetry.  In the eight band model for cubic semiconductors $g = g_0
- g_r - 2E_p\Delta/3(E_g-V)( E_g - V+ \Delta)$, where $E_p$ is Kane
energy, $E_g$ is a band gap and $g_r$ is a correction from remote
bands, and the effective Rashba constant is $\alpha_R \propto \left.
\partial g/\partial x \right|_{x=0} \propto \Delta$.  Using the
expression for $g$ in the 14 band model one finds that the correction
to $\alpha_R$ is proportional to $\Delta^c$.

Another useful example is provided by the degenerate valence band at
the $\Gamma$ point described by the envelope Hamiltonians obtained in
\cite{lut}. 
  We consider 
two cases of $\Delta=0$ and $\Delta
\rightarrow \infty $.  The remarkable property of the respective
envelope Lagrangians obtained according to Eq.  (\ref{hl}) is the
existence of the asymmetric term with ${\bm K}(\Gamma_{15})
=-i(1+3\kappa)/m_0{\bm I}$ even in the case $\Delta=0$:
\begin{eqnarray}
 {\cal L}_{\rm SIA}(\Gamma_{15}) =
-\frac{i\hbar^2}{2m_0}(1+3\kappa) {\bm \nabla} \Psi_\alpha^* [ {\bm I}
\times {\bm \nabla} \Psi_\alpha] \, . 
 \end{eqnarray} 
 Here $\kappa $
is the magnetic Luttinger constant, $\hat {\bm I}$ is the internal
angular momentum operator ($I=1$) and $\Psi_\alpha$, $\alpha=0,\pm 1$,
is the 3 component envelope function.  The SIA component of the
velocity operator $\hat {\bm V}_{\rm
so}={i\hbar}(1+3\kappa)/{m_0}[{\bm I} \times {\bm k}]$ induces a new
short range SIA term in the heterostructure Hamiltonian for the
$\Gamma_{15}$ holes as well as the ${\bm I}$-dependent boundary
conditions.  This leads to the splitting of the heavy hole subband in
asymmetric structures mediated by the interaction with light hole
states.  Note that it is this asymmetric term ${\cal L}_{\rm
SIA}(\Gamma_{15})$ which is responsible for the magnetic energy term
$\propto \mu_B(1+3\kappa) ({\bm I} {\bm H})$ in the bulk Hamiltonian
of Ref.  \cite{lut}.

In the case of $\Delta \rightarrow \infty $, the top of valence band
is four-fold degenerate corresponding to the $J=3/2$ subspace of the
total internal momentum ${\bm J}={\bm I} + 1/2{\bm \sigma}$.  We
obtain the asymmetric term in the envelope Lagrangian with ${\bm
K}(\Gamma_{8})=-i(2/3+2\kappa)/m_0{\bm J} - iq/m_0{\bm F}$, where $q$
is cubically anisotropic magnetic Luttinger constant and ${\bm
F}=(F_x,F_y,F_z) \equiv (J_x^3,J_y^3,J_z^3)$: 
 \begin{eqnarray}{\cal
L}_{\rm SIA}(\Gamma_{8}) = \frac{\hbar^2}{2} {\bm \nabla}
\Psi_\alpha^* [ {\bm K}(\Gamma_{8}) \times {\bm \nabla} \Psi_\alpha]
\, . 
 \label{sia8} \end{eqnarray}
  Here $\Psi_\alpha$, $\alpha=\pm
3/2,\pm 1/2$, is the 4 component envelope wave function.  The SIA
component of the velocity operator $\hat {\bm V}_{\rm
so}={i\hbar}/m_0((2/3+2\kappa)[{\bm J} \times {\bm k}] +q[{\bm F}
\times {\bm k}])$ produces a new short range SIA term in the
heterostructure Hamiltonian as well as the asymmetric contribution to
the boundary conditions of the $\Gamma_{8}$ holes (see
\cite{pfeffer,foreman}).  The very same asymmetric term ${\cal L}_{\rm
SIA}(\Gamma_{8})$ induces the magnetic energy terms $\propto ({\bm J}
{\bm H})$ and $\propto q ({\bm F} {\bm H})$ in the bulk Hamiltonian of
Ref.  \cite{lut} as well as the macroscopic SIA term $H^{\rm so}_{8v}=
\beta_1[{\bm J} \times {\bm k}]{\bm E} +\beta_2[{\bm F} \times {\bm
k}]{\bm E}$ postulated recently in Ref.  \cite{winkler}.  The
cubically anisotropic constant $q$ and, consequently, $\beta_2$ are
proportional to $\Delta^c$ and usually small.  Considering the
dependence of $\kappa=\kappa_r + E_p/6(E_g+V)$ on $V = -|e|{\cal E
}x$, where $ \kappa_r$ is the contribution from remote bands, we are
able to derive for the first time the effective SIA constant $\beta_1$
for the $\Gamma_8$ valence band: 
 \begin{equation}
\beta_1=\frac{|e|\hbar^2}{6m_0}\frac{E_p}{E_g^2} \, .  
\end{equation}
We found that in contrast to the Rashba constant, $\beta_1$ is not
proportional to the spin--orbit splittings $\Delta$ or $\Delta_c$.

The short--range SIA contribution to the zero field splitting becomes
more pronounced in the case of the interface between very dissimilar
materials.  The general Hamiltonian $\hat H_{AB}$ of Eqs.
(\ref{hab},\ref{hsur}) enables to describe, for example, the interface
coupling between $\Gamma_6$ electrons and $\Gamma_8$ holes and the SIA
effects in type II and type III quantum wells, and 
to take into consideration the microscopic
asymmetry 
 caused by the nonequivalence of two
opposite interfaces (see \cite{as}).

Finally, the developed approach reveals a new short--range magnetic
energy term in the heterostructure Hamiltonian $\hat H_{AB}$ in the
presence of the in-plane external magnetic field.  Indeed, choosing
${\bm H} \parallel {\bm z}$ we obtain the new term in the velocity
operators $V_\tau \Psi$ proportional to $Hx(\hat K_z \Psi)$.  In small
fields this term can be treated as a perturbation.  The short-range
contribution to the Zeeman energy is proportional to the sum of the
discontinuities of $(\Psi^*\hat K_z \Psi)$ at the interfaces.  Unlike
the zero field splitting, this interface magnetic energy contribution
is present even in completely symmetric $2D$ structures as well as in
spherical dots (see \cite{so}).

In conclusion, the variational least action principle allows to
consistently extend the envelope function approach to the
 heterostructures with abrupt interfaces.  
 The SRI terms in the
heterostructure Hamiltonian $\hat H_{AB}$ and the GBC  
are equally valid representation of the interface properties and
can be written for
any interface between dissimilar materials including the case $N_A \ne
N_B$.  For lattice-matched heterostructures ($N_A=N_B=N$) the obtained
$\hat H_{AB}$ allows direct comparison with previously derived
Hamiltonians \cite{foremanburt,volkov,foreman}.  The discretization of
$\hat H_{AB}$ for numerical calculations is straightforward and
requires no additional symmetrization.  All macroscopic and
short-range interface SIA terms as well as the magnetic energy terms
in $\hat H_{AB}$ originate from the asymmetric term ${\cal L}_{\rm
SIA}$ in the bulk envelope Lagrangian.

We are grateful to R.  Suris for attracting our attention to the importance of
the variational principle for boundary condition problems.  A.V.  Rodina
acknowledges the support from the Swiss National Science Foundation.

\end{document}